\documentclass[11pt,a4paper,preprint]{article}
\pdfoutput=1
\RequirePackage{lineno}
\usepackage{jcappub}
\usepackage{graphicx,psfrag}

\begin{document}

\newcommand{\beq}{\begin{equation}}
\newcommand{\eeq}{\end{equation}}
\newcommand{\bear}{\begin{eqnarray}}
\newcommand{\eear}{\end{eqnarray}} \newcommand{\ba}{\begin{array}}
\newcommand{\ea}{\end{array}}
\newcommand{\lae}{\begin{array}{c}\,\sim\vspace{-1.7em}\\< 
\end{array}}
\newcommand{\gae}{\begin{array}{c}\,\sim\vspace{-1.7em}\\> 
\end{array}}


\title{Constraints on  Self Interacting Dark Matter from IceCube Results} 


\author[a]{Ivone F.~M.~Albuquerque,} 
\author[b]{Carlos P\'erez~de los Heros}
\author[a]{and Denis S. Robertson} 

\affiliation[a]{Instituto de F\'isica, Universidade de S\~ao Paulo, S\~ao Paulo, Brazil}
\affiliation[b]{Department of Physics and Astronomy. Uppsala University. Uppsala. Sweden}

\emailAdd{ifreire@if.usp.br}
\emailAdd{denistefanrs@gmail.com}
\emailAdd{cph@physics.uu.se}

\date{\today}

\abstract{
If dark matter particles self-interact, their capture by astrophysical
objects should be enhanced. As a
  consequence, the rate by which they annihilate at the center of the
  object will increase.   
  If their self scattering is strong, it can be observed indirectly through an
  enhancement of the flux of their annihilation products. Here we investigate the effect of self-interaction on the
  neutrino flux produced by annihilating dark matter in the center of
  the Sun.  We consider annihilation
  into two channels: $W^+W^-$ (or $\tau^+\tau^-$ for a dark matter
  mass below the $W$ mass) and $b\overline{b}$. 
  We estimate the event rate in the IceCube detector,
  using its 79-string configuration, and compare our prediction
  to their experimental results, hence probing dark matter self
  interacting models.}


\maketitle

\section{\label{introduction} \bf Introduction}
Current cosmological observations indicate that our Universe
is flat, and composed mainly by dark matter and dark
energy. Observations at large scales are very well fit by
collisionless cold dark 
matter models (CDM). However these models present potential
problems related to small scale structure formation, one of them being
referred to as the core/cusp problem~\cite{blok}. While 
structure formation simulations~\cite{benm,flores,nfw}, based on CDM
models, present a steep cusp density profile,
observations of dwarf galaxies~\cite{carig,lake,jobin,walker,oh}
indicate a cored density profile rather than a cusped one. Also 
CDM simulations evolve to very dense subhaloes of Milky Way type galaxies, which is
a problem since these cannot host its brightest satellites
\cite{toobig,bbk}.  This discrepancy is known as the ``too big to
  fail problem'', given that it would be hard to miss the observation of
  these substructures.

Warm dark matter models (WDM) have been proposed as a solution to these
inconsistencies, since  it was expected that they should develop shallower
density profiles at small radius, and would avoid unreasonably dense 
subhaloes~\cite{lovell}. However there are also discrepancies
between simulations and observations, where the core/cusp problem is
not solved by WDM~\cite{maccio}, and thermal WDM candidates seem to be ruled
out by Lyman-$\alpha$ forest results~\cite{lyma}.

Another very interesting solution to these small scale problems is to
consider dark matter self-interactions (SIDM) \cite{spst}, where dark matter
particles scatter among themselves,  instead of collisionless CDM models. If this scatter is strong
enough, the halo central density profile will be softened in relation
to a pure CDM model. Cosmological simulations~\cite{irv1,irv2} show that SIDM models
with a ratio between the self-interaction cross sections over the
dark matter mass, $\sigma_{\chi \chi} / m_\chi \sim \vartheta( 0.1 {\rm cm^2/g})$ will reconcile simulations with the
observed dwarf galaxies properties, while these self-interactions will
not modify the CDM behaviour at large scale. 

It should be noticed however that the small scale potential problems
  might not be associated with the CDM models themselves, but with
  other structure evolution features. As examples, the ``too big to
  fail problem'' can be solved by the inclusion of
  baryons in structure formation simulations which changes the shape
  of dark matter
  profiles \cite{brooks}, or by the fact that similar host halos present variations in their
  subhalos properties \cite{purcell}. The inclusion of baryons, as well as
  three dimensional mass distributions \cite{hayashi} 
  might also contribute to cored profiles \cite{governato}. Although
  these solutions might bring CDM simulations to agree with
  observations, the SIDM solution should also be contemplated and seriously explored.

In this paper we probe SIDM models through the neutrino flux produced
from dark matter annihilation in
the center of the Sun. Dark matter scattering off the dark matter that has
already been captured in the Sun's potential well will enhance its
capture rate, and consequently its annihilation rate. In this way, if dark
matter self-interacts in the Sun, the neutrino flux should be enhanced when compared to the one produced by
collisionless dark matter annihilation. This was noted in \cite{zent}.

Note that, in scenarios where the $\sigma_{\chi \chi}$ is
  velocity-independent,  the annihilation process by itself does not differ
between collisionless CDM and SIDM. The enhancement comes exclusively through
the capture rate. Not only there is an increase in the number of dark matter
particles that are captured, but as the equilibrium between capture and
annihilation happens faster when considering self-interactions, the
maximum annihilation rate is reached earlier than in pure CDM models.

The most robust constraint to SIDM comes from an analysis of the
Bullet Cluster matter distribution~\cite{bullet}, which excludes
$\sigma_{\chi \chi} / m_\chi > 1.25 \;{\rm cm^2/g}$. There are also
constraints from an analysis on the core
  densities of galaxy clusters, low mass spiral galaxies and dwarf spheroidal galaxies
  \cite{irv1}, and from halo shapes~\cite{irv2}, excluding $\sigma_{\chi \chi} /
m_\chi > 1.0 \;{\rm cm^2/g}$. Another interesting analysis,
based on the kinematics of dwarf
spheroidals~\cite{zav}, estimates that SIDM will only alleviate CDM
small scale problems when  $\sigma_{\chi \chi} / m_\chi > 0.1 \;{\rm
  cm^2/g}$. These analyses are more assumption dependent than the
Bullet Cluster analysis. If we
take all these limits as robust, and consider the area where SIDM
  is expected to be effective, there is still a small,
but non-the-less interesting, non probed region between  $(1.0 > \sigma_{\chi \chi} /
m_\chi > 0.1) \;{\rm cm^2/g}$. Our analysis will probe most of this 
allowed region. At the same time it will, in an
independent way, probe the parameter space region excluded  by the
Bullet Cluster and halo shapes analyses.

We note that these analyses imply that only very strong $\sigma_{\chi\chi}$, at the
$\vartheta(10^{-22}$~cm$^2$), can solve CDM potential cosmological
problems. Although these are stronger than cross sections for
nucleon-nucleon interactions, and not at first hand expected, it is important
to probe these allowed regions of SIDM.  

Our investigation proceeds by computing the neutrino flux from dark matter
annihilations in the Sun through Monte Carlo simulations, and we consider
two extreme cases as benchmarks:  annihilation 
into $W^+W^-$ (or $\tau^+\tau^-$
when the dark matter mass is less than the $W$ mass) and
$b\overline{b}$. We determine the expected 
neutrino flux at the IceCube detector, and based on the fact that there was no
measured anomalous neutrino flux from the Sun \cite{icprl} we set limits on the
$(\sigma_{\chi\chi},m_\chi)$ parameter space.

In the next section we discuss the dark matter capture and annihilation rates
enhancement due to self-interactions. We then describe how we determine the neutrino
rate in IceCube. Following we analyze our results, and compare our predictions to IceCube
results and finally describe our conclusions. 

\section{\bf \label{cptann} Dark Matter Capture and Annihilation Rates Enhancement
  due to Self-Interactions}

If dark matter self interacts, the evolution of its number
($N_\chi$) in the Sun will follow

\beq
 \dot{N_\chi} \;=\; \Gamma_C \,+\, \Gamma_{\chi\chi} \,-\, \Gamma_A, 
\label{eq:dmevo}
\eeq
where $\Gamma_C$ and $\Gamma_{\chi\chi}$  are the capture rates
for dark matter particles that
interact with the Sun's nuclei, and due to self-interactions, respectively. 
$\Gamma_A$ is the annihilation rate, which equals
$N_\chi^2 C_A$, where $C_A$
depends on the dark matter distribution in the Sun~\cite{gould,jkg}, and is
given by $C_A\,\equiv \, \left< \sigma_A v \right>/V_{\text{eff}}$, where
  $\left< \sigma_A v \right>$ is the relative velocity averaged
  annihilation cross section. $V_{\text{eff}}$ represents the dark matter effective volume at the
center of the Sun. Assuming an isothermal distribution, and taking the Sun's temperature as $1.57 \times 10^7$~K
\cite{tsun},  $V_{\text{eff}} = 6.9 \times 10^{27} \times (100 \,{\rm GeV}/m_\chi)^{3/2}$~cm$^3$.

The dark matter capture rate due to elastic scatter off the Sun's nuclei is given
by~\cite{gould,jkg}

\beq
\Gamma_C^{SI} \;=\; 6.8 \times 10^{20} {\rm s^{-1}}\, \frac{n_\chi}{0.4 {\rm cm^{-3}}}\, \frac{\sigma_{\chi n}}{10^{-44}{\rm cm^2}}
\,\frac{270 {\rm km/s}}{\overline{v}} \, \sum F_i(m_\chi) \, A_i^3 \,
\left( \frac{m_\chi + m_p}{m_\chi + m_{N_i}} \right)^2 \, \delta_i\,
K \left( \frac{m_\chi}{m_{N_i}} \right)^2,
\label{eq:cpt}
\eeq
and depends on the dark matter local number density
$n_\chi$, mass, velocity dispersion $\overline{v}$, spin
independent dark matter nucleon cross
section $\sigma_{\chi n}$, and the sum is over all Sun's nuclear species i.
$F_i(m_\chi)$ accounts for the interaction form factor suppression,
$\delta_i$ for the mass fraction and distribution of the various
nuclei over the Sun, and $K(\frac{m_\chi}{m_{N_i}})$ is a kinematic suppression factor.
$m_{N_i}$ is the nucleus $i$ mass and $A_i$ its atomic mass. The cross sections for dark matter-nuclei
interaction $\sigma_{\chi N_i}$ relate to $\sigma_{\chi n}$ as  $\sigma_{\chi N_i} = \sigma_{\chi n}\,
A_i^2\, \left(\frac{\mu_{\chi N_i}}{\mu_{\chi p}}\right)^2$, where
$\mu$ is the reduced mass, and we approximated $m_{N_i}  \sim A_i m_p$, where $m_p$ is the proton's mass
and is not distinguished from a neutron, in a better
than $1\%$ approximation. To compute $\delta_i$ we take the nucleus
mass fraction as given in \cite{spaces} and its distribution from
\cite{jkg}. The distribution can be conveniently represented by a
dimensionless gravitational potential $v^2_{esc}(r) /
v^2_{esc}(R_\odot)$ as described in \cite{gould}.

When computing the capture rate due
to self-interactions, all terms in the sum of the above equation have
to account for self instead of $\chi$-nuclei
interactions. The number of captured dark matter particles increases with
time, until the capture rate reaches equilibrium with the
annihilation rate. The self-interaction capture rate is given by ~\cite{zent} :

\beq
\label{eq:gcc}
\Gamma_{\chi\chi} \;=\; N_\chi \,\sqrt{\frac{3}{2}} \,n_{\chi}
\,\sigma_{\chi\chi}\,\frac{v_{esc}^2(R_\odot)}{\overline{v}}\,\left<\phi_\chi\right>\,\frac{{\rm
    erf}(\eta)}{\eta}
\eeq
where $\left<\phi_\chi\right>$ accounts for the dark matter distribution, which for
the Sun is
approximately 5.1 \cite{gould2} since it concentrates more towards the center,
$v_{esc}(R_\odot) = 617.5$~km/s is the Sun's escape velocity at its
surface, and $\eta\,=\,
\sqrt{\frac{3}{2}}\,\frac{v_\odot}{\overline{v}}$, where $v_\odot = 220$~km/s is
  the Sun velocity through the dark matter halo, which we assume has a Maxwell-Boltzman
  distribution with a velocity dispersion of
  $\overline{v} = 270$~km/s. We take the local dark matter density 
  as $\rho_\chi\,=\,0.4$~GeV/cm$^3$ \cite{sal,ullio}. It is
    useful to define $\Gamma_{\chi\chi}^{\prime} = \Gamma / N_\chi$,
    noting that it is independent 
    of $N_\chi$.

\begin{figure}[t]
\vspace*{-2.cm}
\hspace*{-1.cm}
\includegraphics[width=1.1\textwidth]{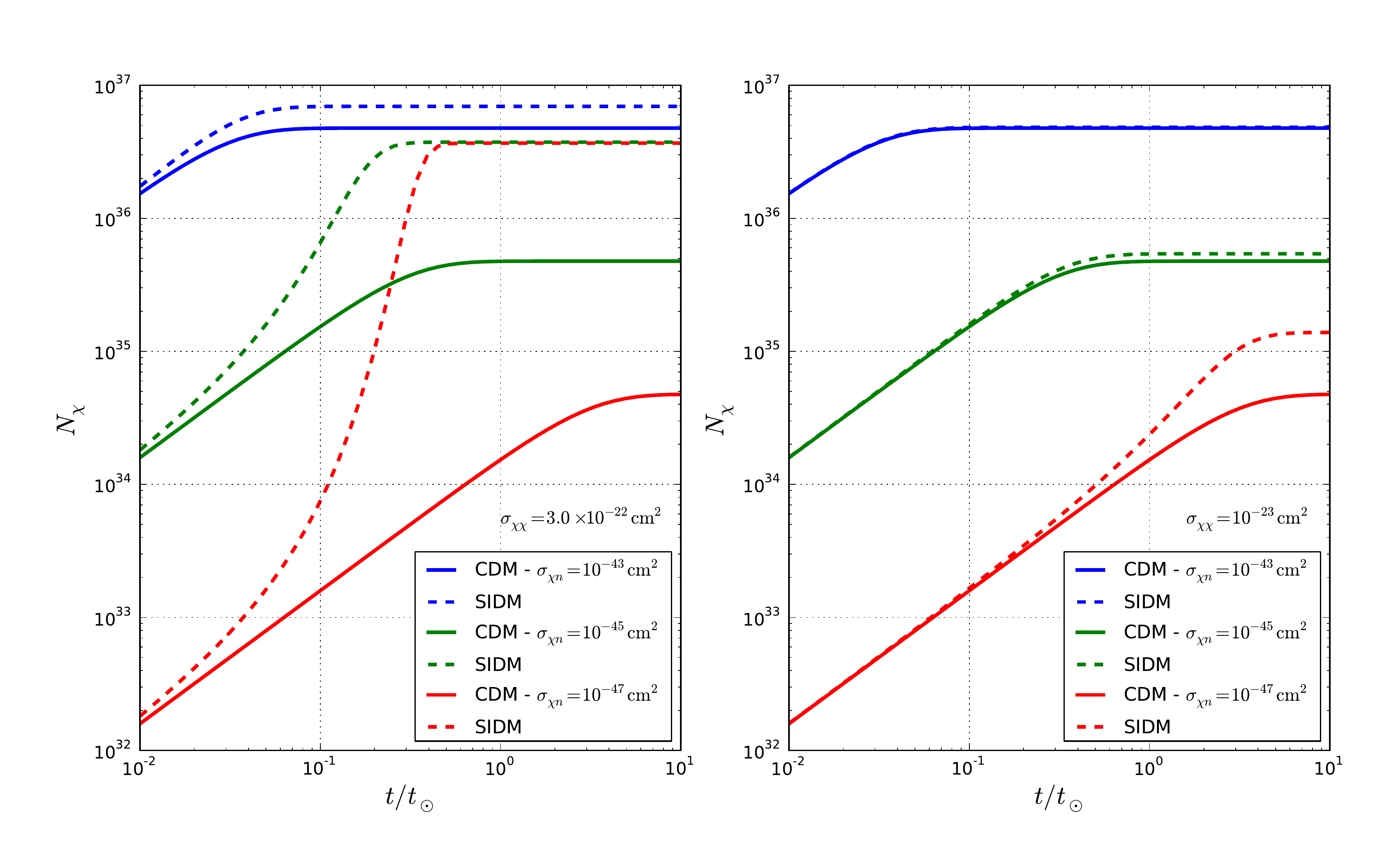}
\vspace*{-.25cm}
\caption{\label{fig:evo} Time evolution of dark matter particles in the Sun. On
the left $\sigma_{\chi\chi} = 3 \times 10^{-22}$~cm$^2$ and on the
right $\sigma_{\chi\chi} = 10^{-23}$~cm$^2$. Solid curves are for
CDM models while dashed curves for CDM+SIDM. Curves are for different
$\sigma_{\chi n}$ values as labelled.}
\end{figure}

An important effect of SIDM is that the timescale for capture and
annihilation equilibrium, given by
\beq
\tau_{\chi\chi} = \left(\Gamma_C C_A \,+\,
  \frac{\Gamma_{\chi\chi}^{\prime 2}}{4}
\right)^{-1/2},
\eeq
is shorter than for CDM-only models.
This effect can be seen in Figure~\ref{fig:evo}, where we compare the time evolution of the number
of collisionless dark matter particles (CDM models) to CDM+SIDM (SIDM)
models. As in the CDM case, once the equilibrium is reached, 
the annihilation rate is maximum and the number of dark matter particles in the
Sun is stable. The stronger $\sigma_{\chi\chi}$ is, the faster the
equilibrium will be reached. For $\sigma_{\chi\chi} = 3 \times
10^{-22}$~cm$^2$ and a low $\sigma_{\chi n}$ the equilibrium will only
have been reached in the Sun if SIDM is considered.

Figure~\ref{fig:evo} also shows that SIDM models with strong
$\sigma_{\chi\chi} \sim \vartheta(10^{-22}$~cm$^2$), and in the
$\sigma_{\chi n}$ region which is not excluded by direct detection experiments, 
will enhance the flux of dark matter annihilation products from the
center of the Sun. This enhancement is significant when compared to
pure CDM models. One interesting feature to be observed
  \cite{zent} when determining the neutrino flux from dark
  matter annihilation, is that when $\sigma_{\chi \chi}$ is strong
  enough it will become independent of $\sigma_{\chi n}$. This can be
  seen in the left plot of Figure~\ref{fig:evo}, through the convergence
  of the red and green lines.

The Sun's spin-dependent (SD) capture rate will mainly include interactions with
hydrogen. In this case there is no significant form factor
suppression, and by modifying equation \ref{eq:cpt} accordingly, 
the spin-dependent rate is given by:

\beq
\label{eq:gcsd}
\Gamma_C^{SD} \;=\; 17.3 \, \times \, 10^{20} \,{\rm s^{-1}}\, \frac{n_\chi}{0.4
  {\rm cm^{-3}}}\,\frac{270 {\rm km/s}}{\overline{v}}\, \frac{\sigma_{\chi H}}{10^{-44}{\rm cm^2}}
\times K\left(\frac{m_\chi}{m_{H}}\right)
\eeq

The solution for the dark matter number evolution equation (Equation
\ref{eq:dmevo}) is given by

\beq
N_\chi \;=\; \frac{\Gamma_C \tanh{ \left(t / \tau_{\chi
      \chi}\right)}}{\tau_{\chi\chi}^{-1}  - \Gamma_{\chi \chi}^{\prime} \tanh(t /
    \tau_{\chi\chi}) / 2}
\eeq
and allows us to determine the annihilation rate $\Gamma_A = C_A N_\chi^2 / 2$. 

\section{\label{sec:nuflx} \bf Neutrino Flux From SIDM Models}

In order to determine the neutrino flux from dark matter annihilation in the
center of the Sun, we perform Monte Carlo simulations using the
WimpSim package~\cite{wimps}. We simulate a generic dark matter particle $\chi$ and
antiparticle $\overline{\chi}$ annihilating into $W^+W^-$ (or
into $\tau^+\tau^-$ for $m_\chi < m_W$), and to $b\overline{b}$. These
channels were chosen since their decay chain will produce 
neutrinos within a wide energy range. They are also the ones analyzed by the IceCube collaboration, and
therefore allow us to compare our results to theirs.
The $W^+W^-$ and $b\overline{b}$ decay chain will produce neutrinos, either as 
primaries or secondaries, which will be propagated to
the position of the IceCube detector at the Earth. As a result of our simulation, two neutrino energy spectra
$\frac{d\Phi_{\nu_\mu}}{dE}$ at the detector are generated, one for $W^+W^- /\tau^+\tau^-$ channel and the
other for $b\overline{b}$.  We simulate two
sets of events, one corresponding to a winter and the other to a
summer period. We follow the same definition of data sets as IceCube,
where the winter is further split into two sets, one composed by low
energy events, with neutrino energies $E_\nu \lae 95$~ GeV (WL), and the other
to high energy events (WH). The summer set (SL) includes only low energy events,
and their observation requires that the neutrino interaction
  occurs inside the DeepCore \cite{deepc} detector, which is
embedded in IceCube. This requirement rejects down-going atmospheric
muons which traverse the detector, faking a possible signal.

Neutrino oscillations as well as charge and neutral current
interactions are considered, and we assume the standard parameters for neutrino
oscillations~\cite{pdg}. This latter effect will be significant for 
neutrinos travelling from the Sun to the Earth, and the initial flavor composition will differ from the one
near the detector. In this analysis we consider only muon neutrinos
arriving at IceCube.

\begin{figure}[t]
\begin{center}
\includegraphics[scale=0.55]{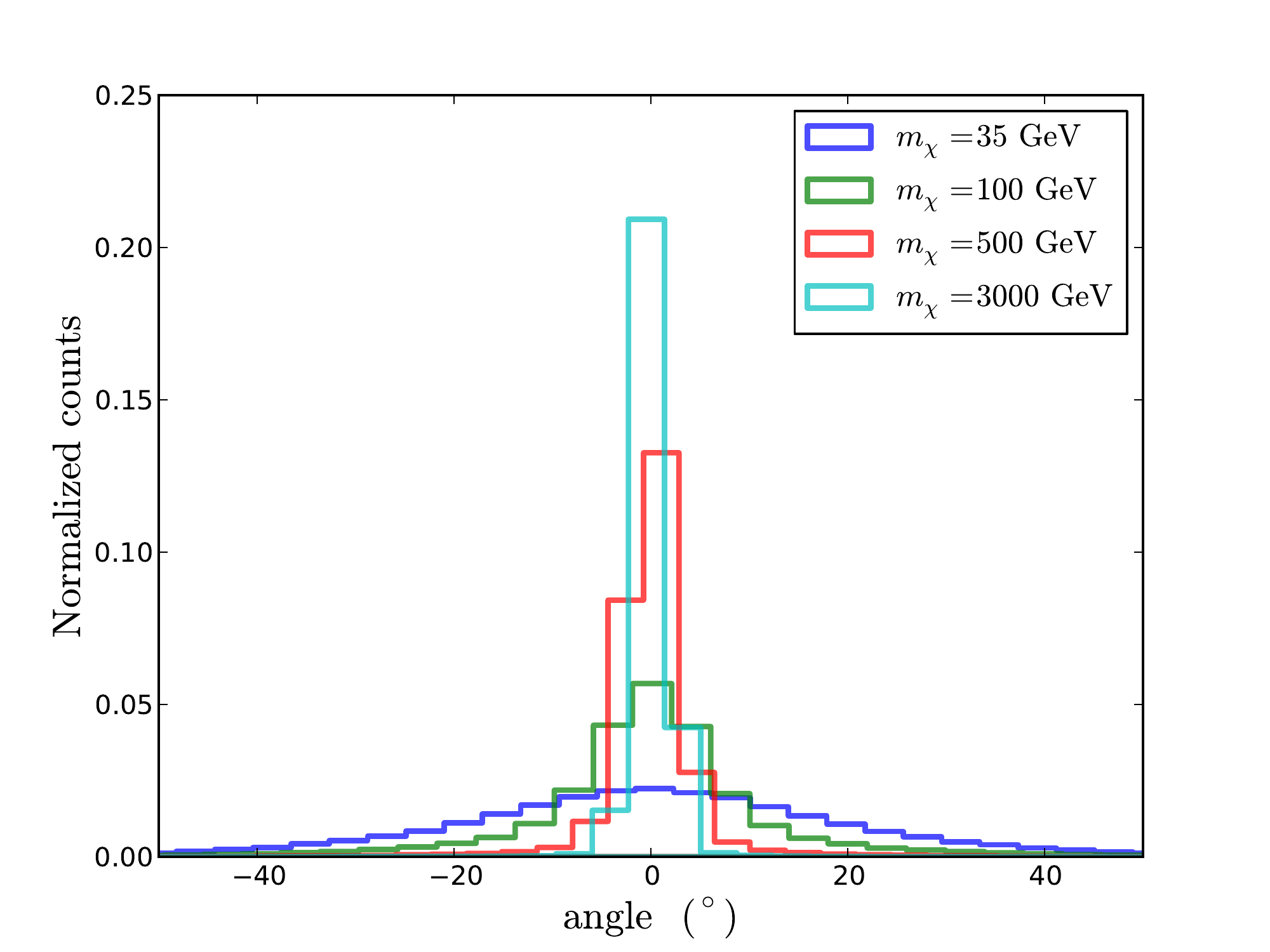}
\vspace*{-.25cm}
\caption{\label{fig:angres} Angular smearing about the Sun-IceCube
  axis due to the experimental angular resolution~\cite{danninger},
  for different dark matter masses. }
\end{center}
\end{figure}

\begin{figure}[t]
\begin{center}
\includegraphics[scale=0.45]{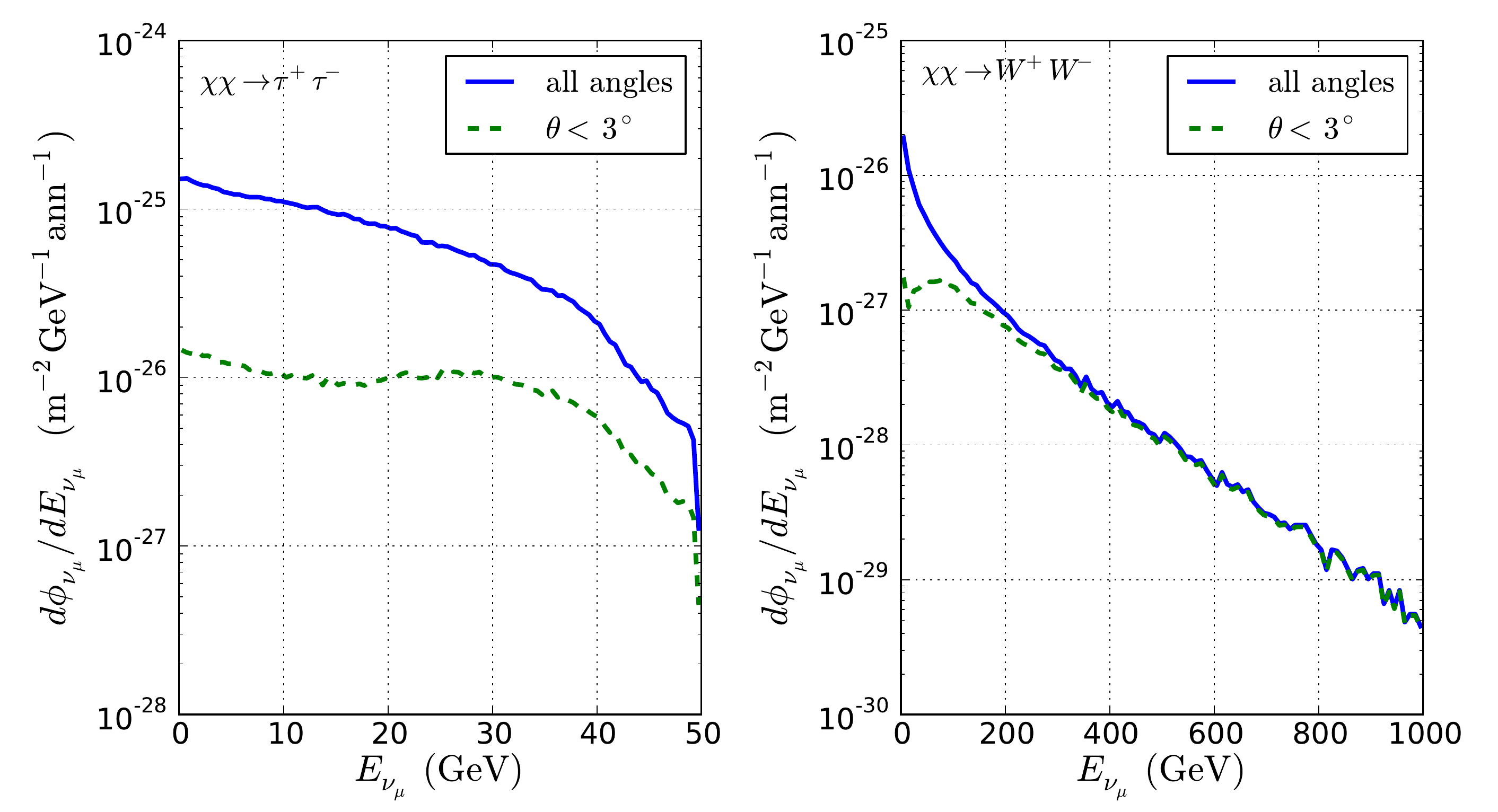}
\vspace*{-.15cm}
\caption{\label{fig:angcut} Neutrino flux versus neutrino energy for
  $m_\chi = 50$~GeV and 1000~GeV annihilating into $W^+W^- / \tau^+\tau^-$. The blue solid line includes all
  simulated events and the green dashed line events with smeared
  angular direction $\theta \protect\lae 3^o$.}
\end{center}
\end{figure}

In order to compare our predictions to observations, we need to
account for IceCube's experimental angular resolution, which is energy
dependent and given in \cite{danninger} . The average angular error is about
$4^o$ for 100 GeV neutrinos, increasing (decreasing) for lower
(higher) energies. We include this reconstruction effect by
smearing the arrival direction of each simulated event by a gaussian
distribution with its $\sigma$ equal to the experimental angular
resolution. Figure~\ref{fig:angres} shows the arrival angular
distributions about the Sun-IceCube axis, for different dark matter
mass values. We then remove all events with a smearing angular
direction  $\theta > 3^o$, which is IceCube's accepted
angular direction for events coming from the Sun~\cite{icprl}. Figure~\ref{fig:angcut} shows
the event reduction due to the angular requirement for the $W^+W^- / \tau^+\tau^-$
channel, for a low (50 GeV) and large (1000 GeV) dark matter mass.

\begin{figure}[t]
\begin{center}
\includegraphics[scale=0.55]{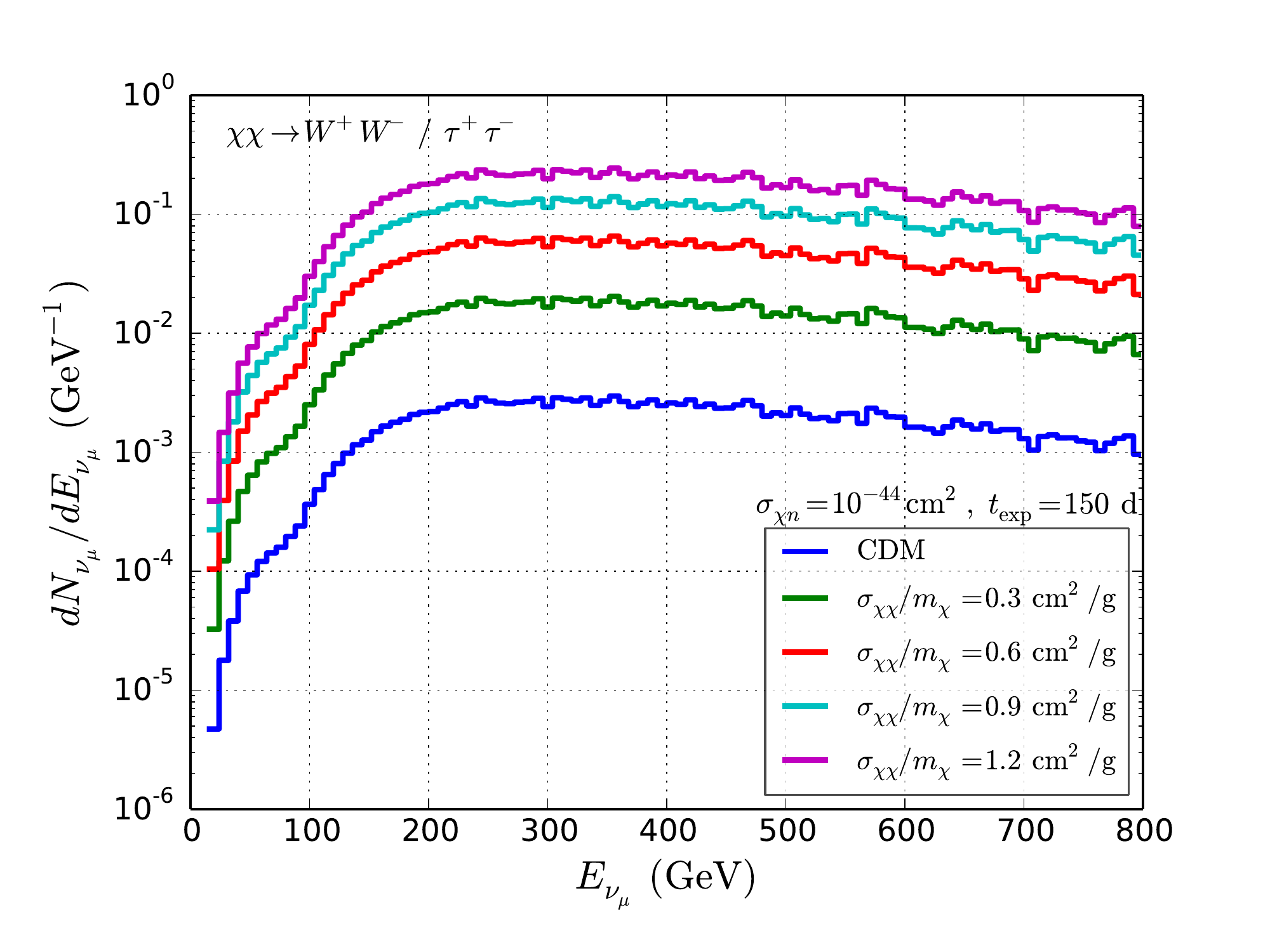}
\vspace*{-.25cm}
\caption{\label{fig:muev} Muon differential energy spectrum at the
  detector as a
  function of the neutrino energy, assuming CDM or CDM+SIDM
  models, for different $\sigma_{\chi\chi} / m_\chi$ values as
  labelled. The integrated number of events is 1.4 for CDM models, 9.6
  (30.8, 66.0, 115.2)  for $\sigma_{\chi\chi} / m_\chi = 0.3 (0.6,
  0.9, 1.2)$~cm$^2$/g. Results are for $m_\chi = 800$~GeV annihilating
into the $W^+W^-$ channel.}
\end{center}
\end{figure}

The expected number of muon neutrinos $N_{\nu_\mu}$, from dark matter
annihilation in the Sun, in IceCube will be given by

\beq 
\label{eq:nev}
  N_{\nu_\mu} = \Gamma_A \; t_\text{exp} \, \int\limits_{E_\text{thr}} 
  \frac{d\Phi_{\nu_\mu}}{dE} \; A_\text{eff}(E) \; 
  \mathrm{d}E,
\eeq
where $t_\text{exp}$ is the exposure time and depends on which data
set is analyzed, being 150 days for the winter 
and 167 days for the summer period. $A_{\text{eff}}$ is IceCube's effective
area, which accounts not only for the energy dependent trigger and analysis efficiencies, but also for the neutrino-nucleon interaction
probability, and the converted muon energy loss before detection. We
take $A_{\text{eff}}$ as given as a function of the neutrino energy for each
data set in \cite{danninger}.
Once we have our prediction for the integrated number of events in IceCube, we can
compare it with the experimental result. Figure~\ref{fig:muev}
exemplifies the predicted spectrum of muons arriving at IceCube, from
800 GeV dark matter annihilation into $W^+W^-$, for different values of $\sigma_{\chi\chi} /
m_\chi$,  and $\sigma_{\chi n} = 1.0 \times
10^{-44}$~cm$^2$ as a function of the neutrino energy. The enhancement
on the expected number of events due to self-interactions is clear in this figure.

\section{\label{exp} \bf Probing SIDM Models}

The IceCube collaboration has searched for signals from dark matter annihilation in
the Sun with its 79-string telescope's configuration~\cite{icprl}. It covered a
large neutrino energy range. By including detection with the DeepCore
array \cite{deepc}, which is embedded in IceCube, they lowered their energy
threshold down to 10 GeV.  IceCube's digital optical modules
that surround this array work as a background veto, allowing  to
discriminate muons produced within the infill, and accumulate useful data
during the summer period. 

Their main background consists of atmospheric
muons and neutrinos. As described in Section~\ref{sec:nuflx} they have three separated data
sets, two for the winter period (WL and WH, where L and H stand for
low and high energies), reflecting the different neutrino energies
covered by both channels analyzed, and one for the low energy
events collected during the summer period (SL). Each of these data
samples, as well as each of the annihilation
channels, were analyzed in different ways given that their background
have different characteristics. IceCube concludes that their data is
consistent with the expected background and impose limits on both WIMP
spin-independent and spin-dependent models \cite{icprl}.

\begin{figure}[t]
\begin{center}
\vspace*{-2.cm}
\hspace*{-1.2cm}
\includegraphics[scale=0.4]{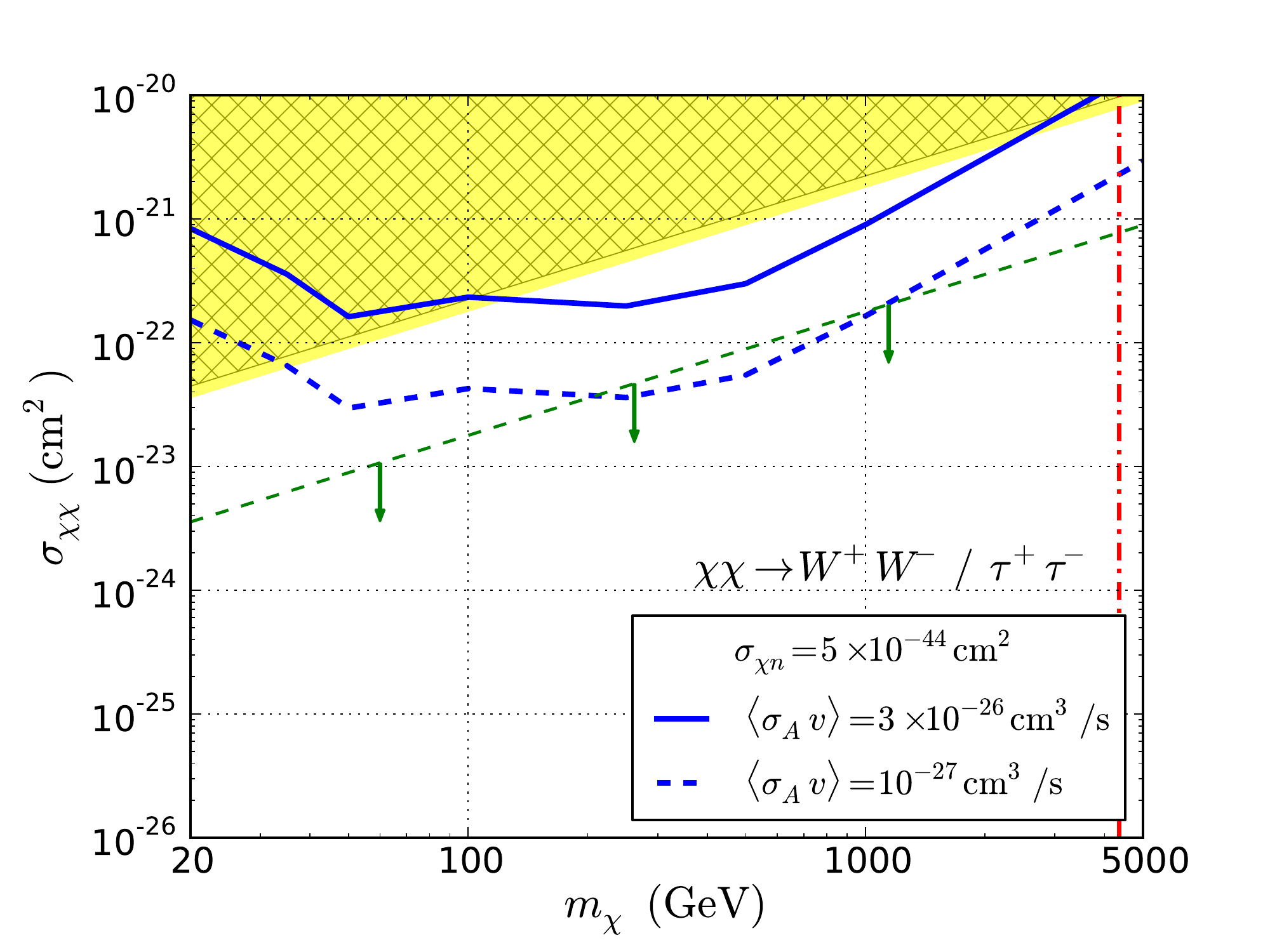}
\includegraphics[scale=0.4]{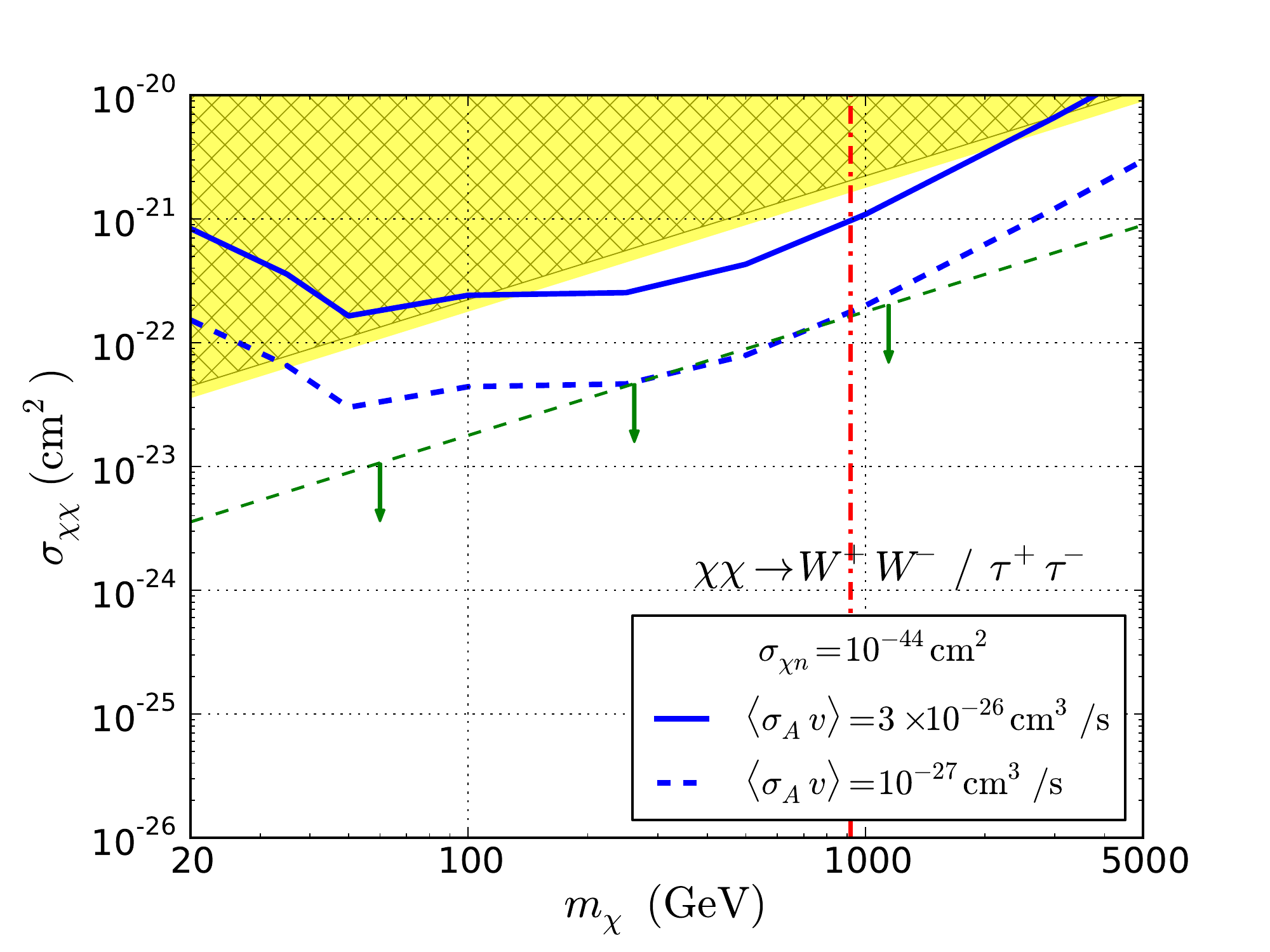}
\hspace*{-1.2cm}
\includegraphics[scale=0.4]{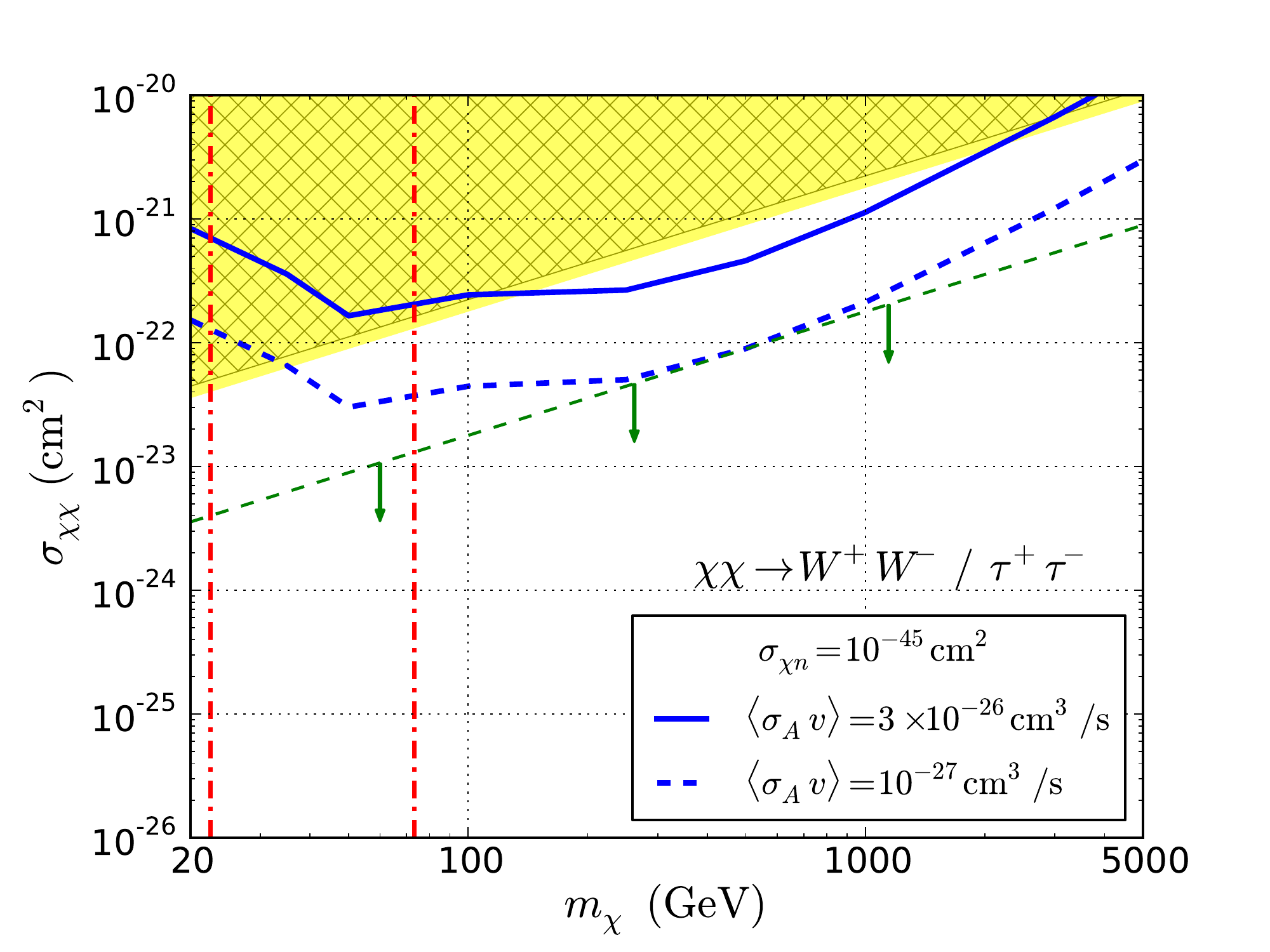}
\includegraphics[scale=0.4]{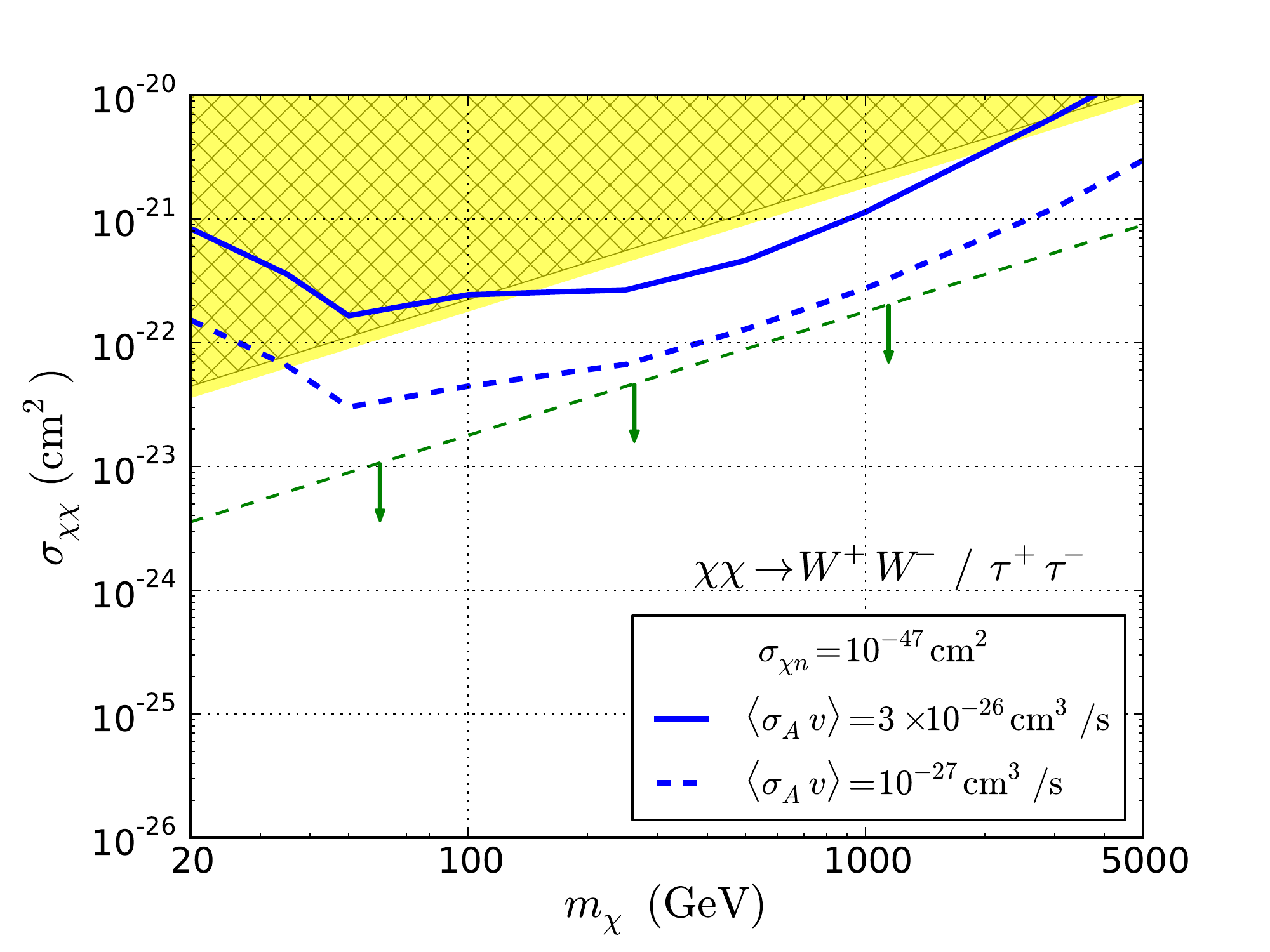}
\vspace*{-.15cm}
\caption{\label{fig:wwexc} Self annihilation cross section
  $\sigma_{\chi\chi}$ versus dark matter mass. The regions above the blue
  curves exclude models with annihilation into $W^+W^- / \tau^+\tau^-$
  at $90\%$~CL by our analysis. The solid (dashed) line is for a
  thermal 
 annihilation cross section $\left<\sigma_A v\right> = 3 (0.1) \times
  10^{-26}$~cm$^3$/s. Each plot considers a different $\sigma_{\chi n}$
  value, as labeled. Exclusion regions from a Bullet Cluster analysis
  \cite{bullet} is shown in black hatches, and by halo shapes \cite{irv2}
  in yellow. The region below the dashed green line, shows the region were
  SIDM is too weak too alleviate CDM potential problems, based on the
  dwarf spheroidals analysis \cite{zav}. The red lines show the direct
detection limits from LUX \cite{lux}, where either the region to the
left or between the lines are excluded. LUX results do
not probe $\sigma_{\chi n} \protect\lae 10^{-47}$~cm$^2$, which is
represented in the bottom right plot.}
\end{center}
\end{figure}
  
IceCube results are expressed as an upper  limit on the number
of signal events at $90\%$~CL ($\mu_{90}$),  as a function of the dark matter
mass and annihilation channel. Any model that predicts a larger
number of events during the data taking period can be ruled out. We use this limit to
compare our predictions and to probe various SIDM models. Our analysis 
probes $\sigma_{\chi\chi}$ as a 
function of $m_\chi$. Figure \ref{fig:wwexc}  shows the regions that are
excluded at $90\%$~CL by our spin-independent $W^+W^- / \tau^+\tau^-$ channel analysis,
where each plot shows limits assuming two values for the  thermal
annihilation cross section ($0.1$ and
$3 \times 10^{-26}$~cm$^3$/s). As can be seen in this figure, a large fraction
of the previously allowed region of the $\sigma_{\chi\chi}$ versus $m_\chi$
parameter space is excluded by our analysis. The smaller (larger) the
$\left<\sigma_A v\right>$ ($\sigma_{\chi n}$) the larger the excluded
region. 

The limits shown in this figure do not change
  significantly for different $\sigma_{\chi n}$ values, as seen from the
  four different plots. This is consistent
  with the fact, mentioned when discussing figure~\ref{fig:evo}, that
  for strong $\sigma_{\chi \chi}$ values, the neutrino flux is
  independent of $\sigma_{\chi n}$.

Figure \ref{fig:wwexc} also shows the regions excluded by analyses of
the Bullet Cluster~\cite{bullet} and the halo shapes
\cite{irv2}, which were briefly discussed in the
  introduction. The region estimated by the dwarf spheroidals analysis~\cite{zav}
as being too weak to solve CDM potential
problems falls below the green line.
Also the region excluded by the direct detection LUX \cite{lux}
collaboration is shown, noting that our analysis
excludes independently SIDM models that fall on
the right hand side of the red line and above the blue lines. LUX does
not probe scenarios with $\sigma_{\chi n} \lae 7 \times 10^{-46}$~cm$^2$.

\begin{figure}[t]
\begin{center}
\vspace*{-2.cm}
\hspace*{-1.2cm}
\includegraphics[scale=0.4]{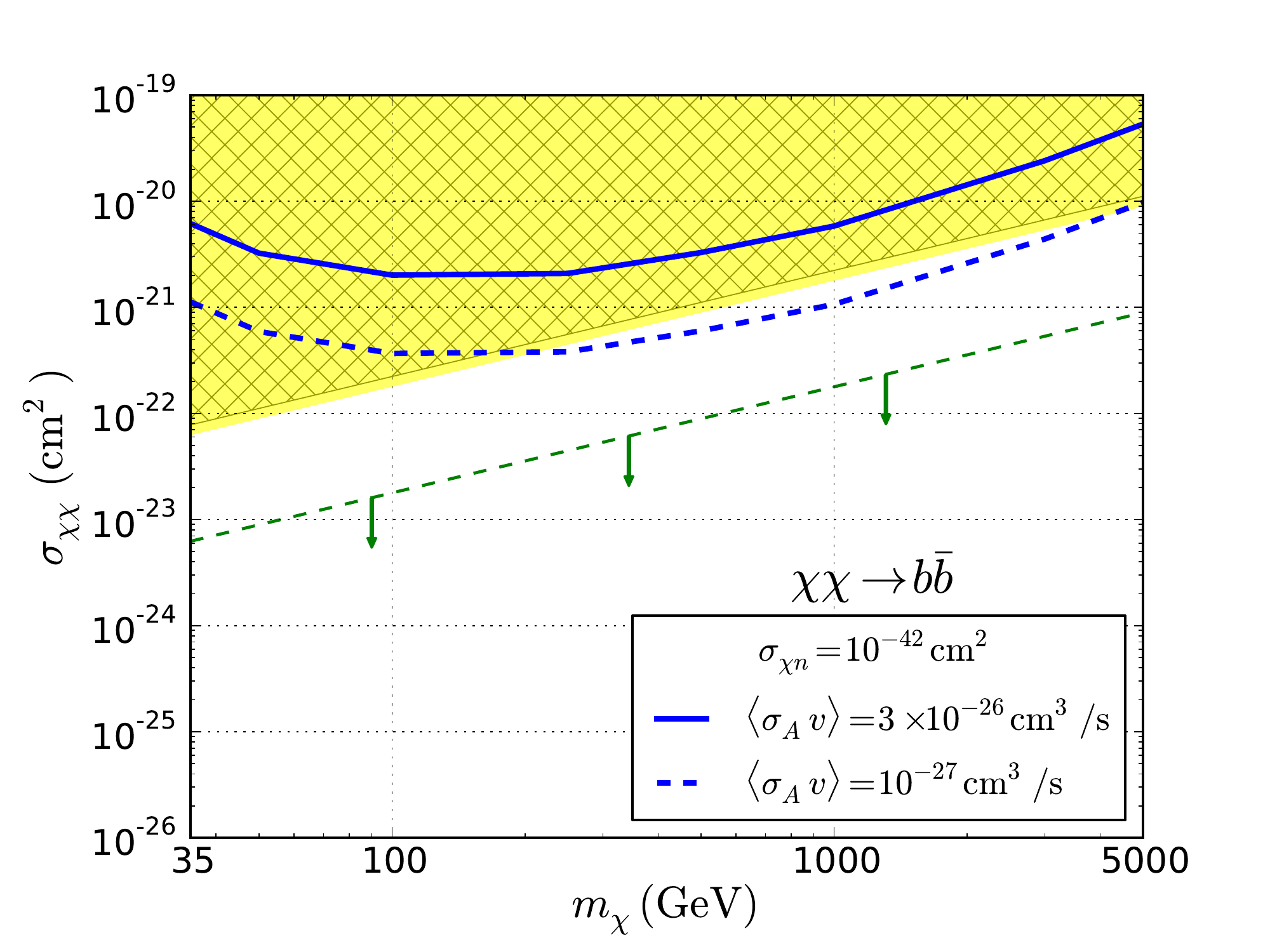}
\includegraphics[scale=0.4]{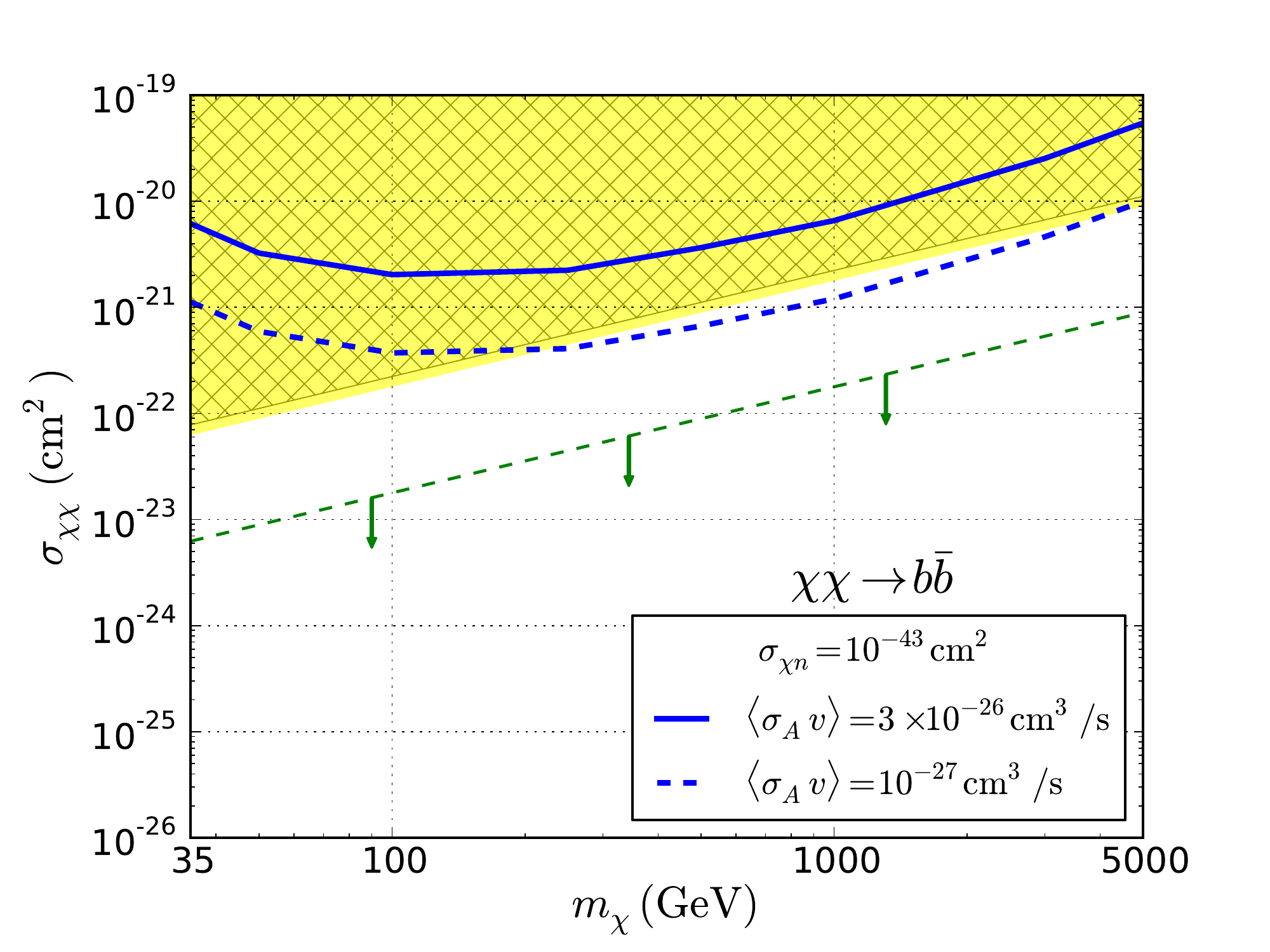}
\hspace*{-1.2cm}
\includegraphics[scale=0.4]{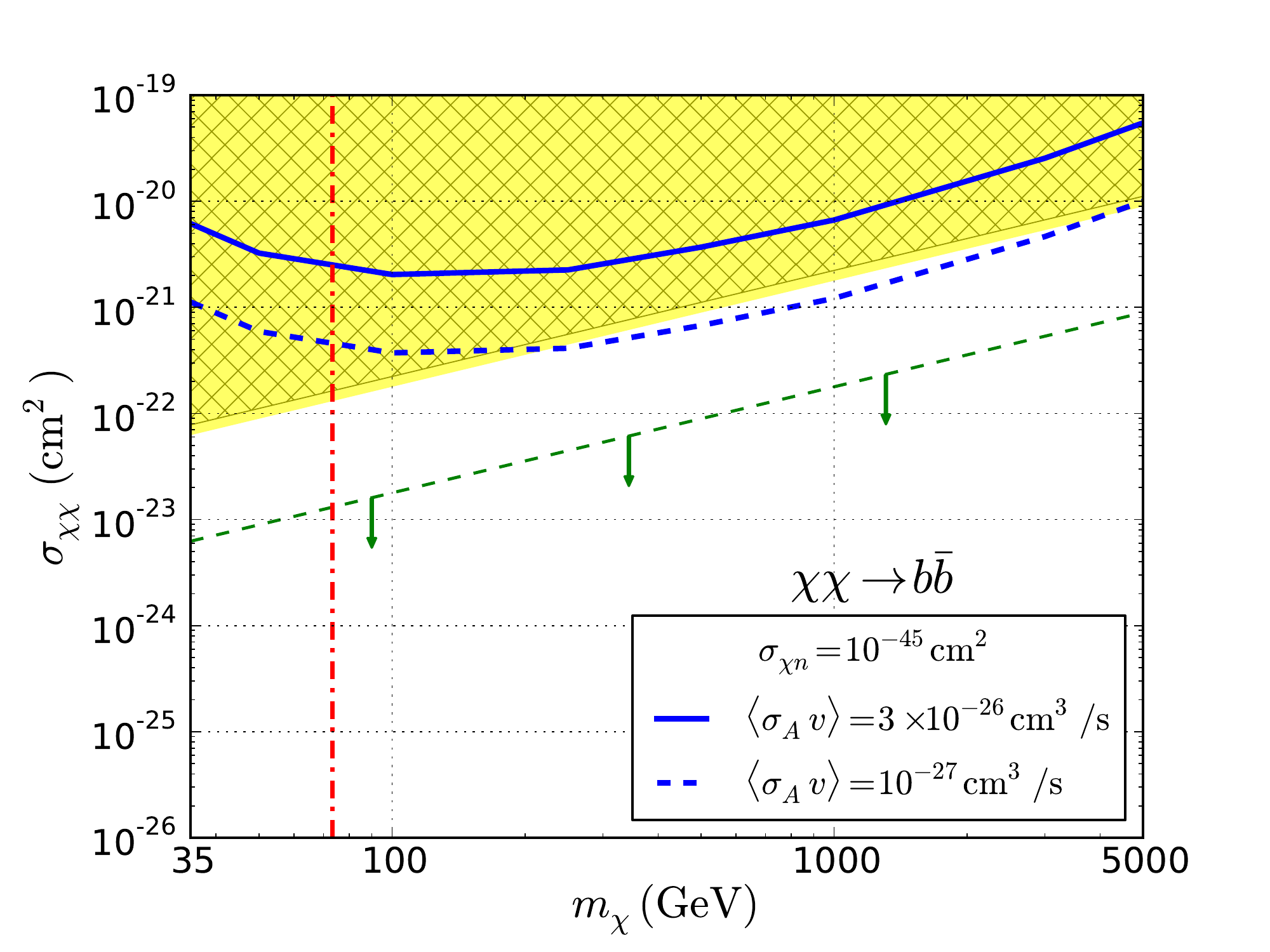}
\includegraphics[scale=0.4]{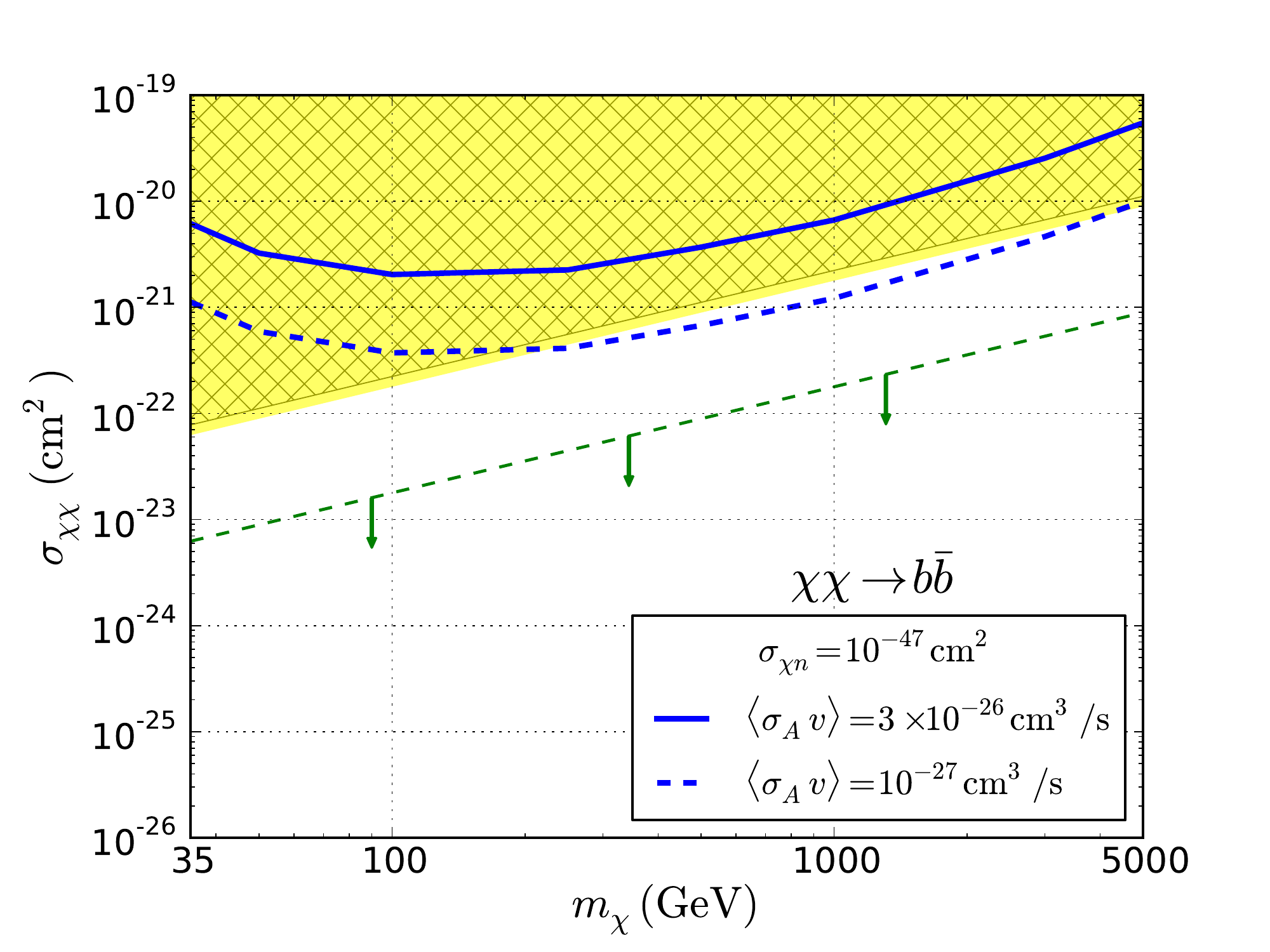}
\vspace*{-.15cm}
\caption{\label{fig:bbexc} Same as previous figure but now for dark matter
  annihilation into $b\overline{b}$. The regions above the blue
  curves are excluded at $90\%$~CL by our analysis. The solid (dashed) line is for a
  thermal annihilation cross section $\left<\sigma_A v\right> = 3 (0.1) \times
  10^{-26}$~cm$^3$/s. Each plot considers a different $\sigma_{\chi
    n}$ value, as labeled.  In these plots
  LUX \cite{lux} direct detection results exclude all the regions
  shown on the top figures, the region to the left of the red lines in the
  leftmost bottom plot, and does not probe the rightmost bottom plot
  region.}
\end{center}
\end{figure}

The $b\overline{b}$ channel results are shown in Figure
\ref{fig:bbexc}. As this channel is softer and produces lower energy neutrinos,
the exclusion regions are smaller than the ones for the $W^+W^- / \tau^+\tau^-$ channel. 
If this annihilation channel holds, it confirms independently most of the region excluded by the
Bullet Cluster and halo shape analyses.

\begin{figure}[t]
\begin{center}
\vspace*{-2.cm}
\hspace*{-1.2cm}
\includegraphics[scale=0.4]{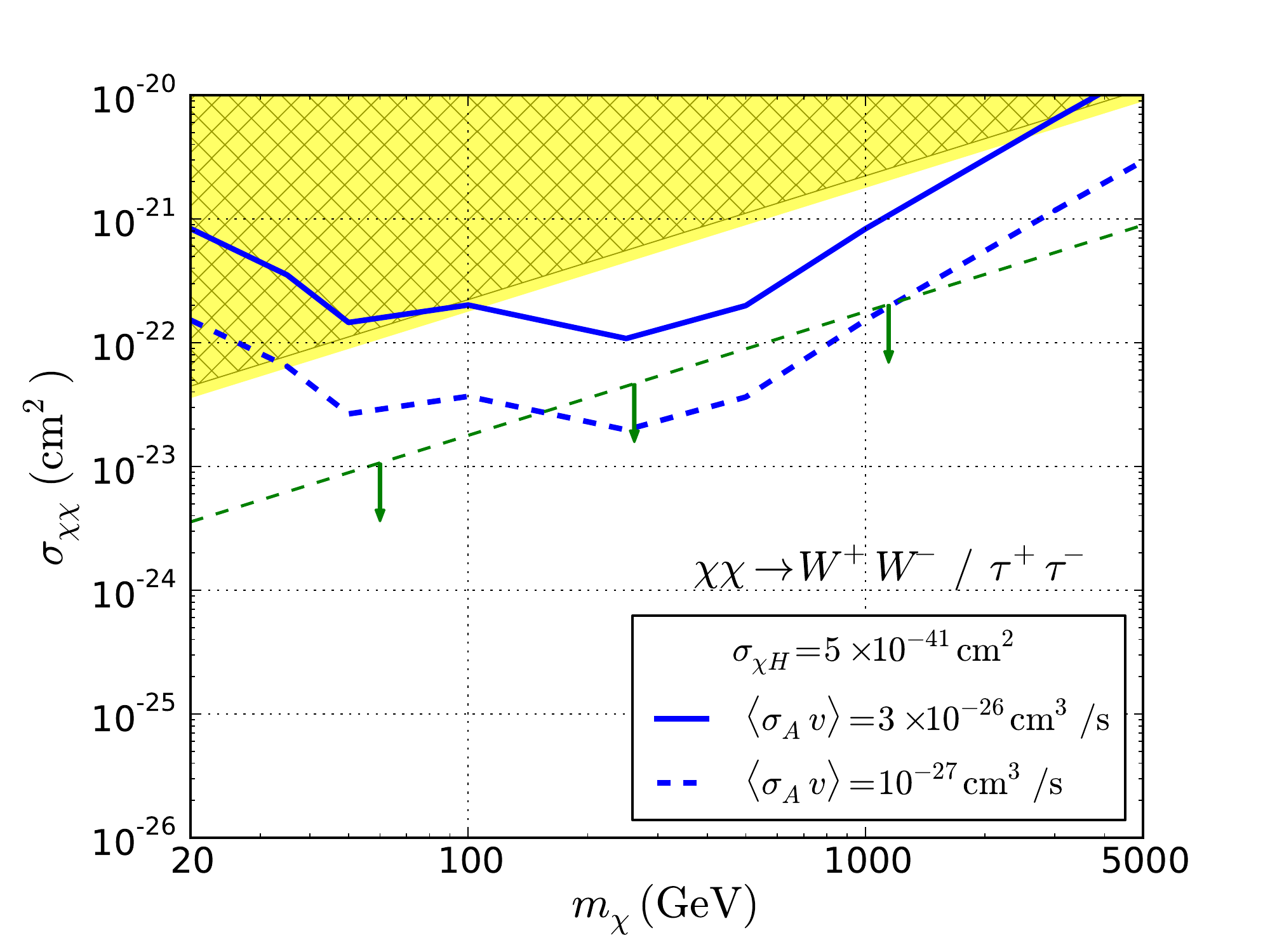}
\includegraphics[scale=0.4]{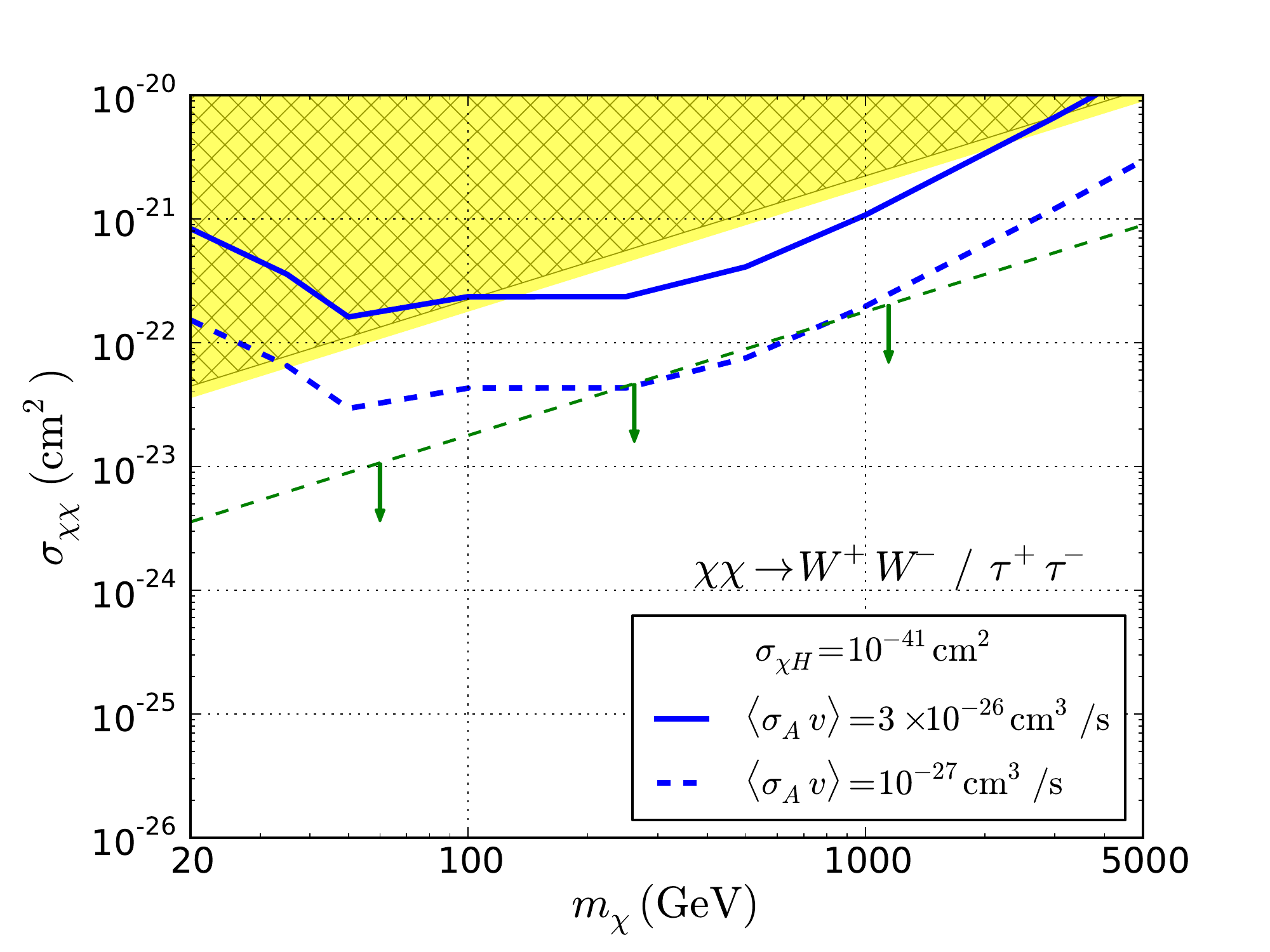}
\hspace*{-1.2cm}
\includegraphics[scale=0.4]{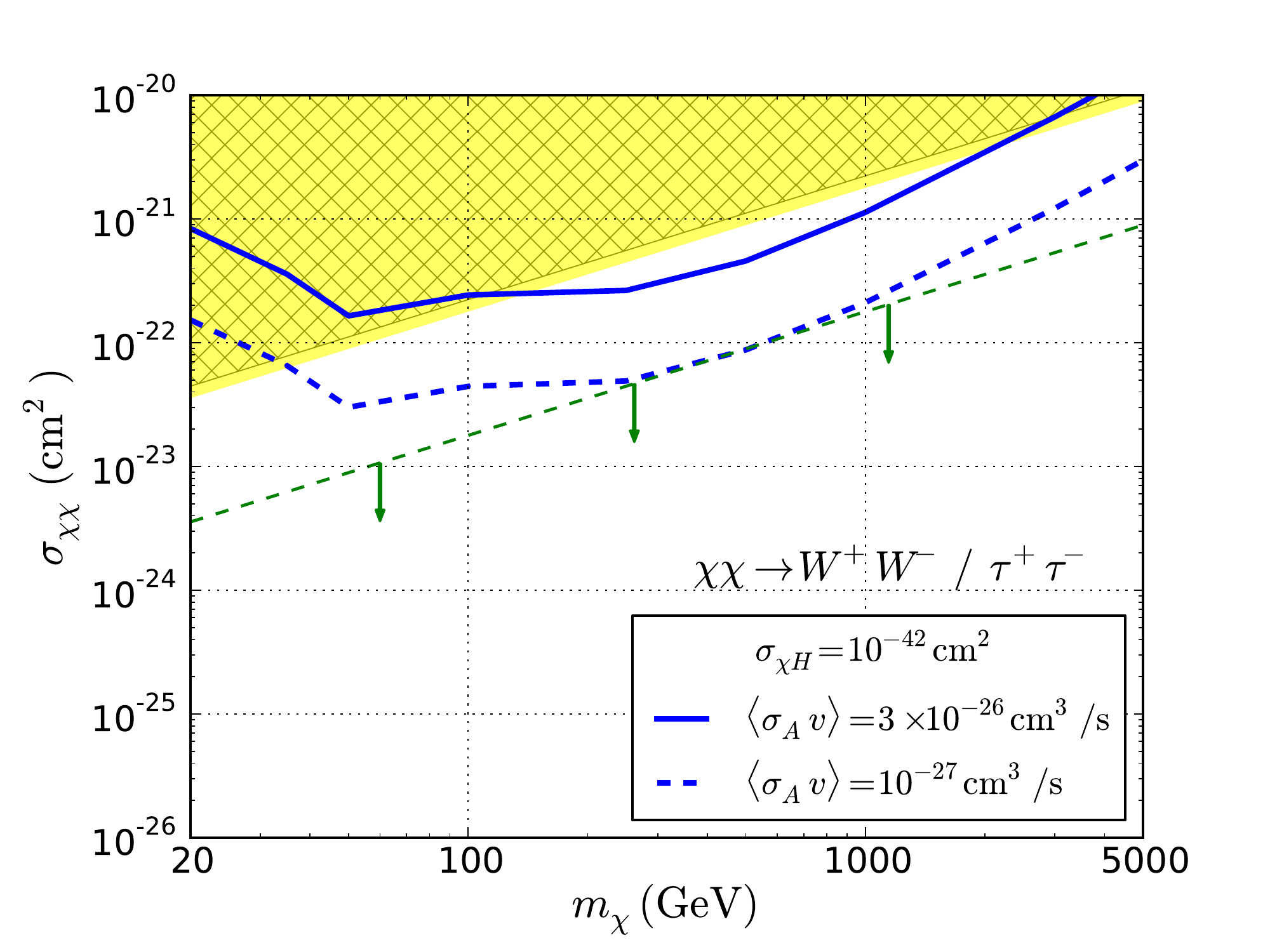}
\includegraphics[scale=0.4]{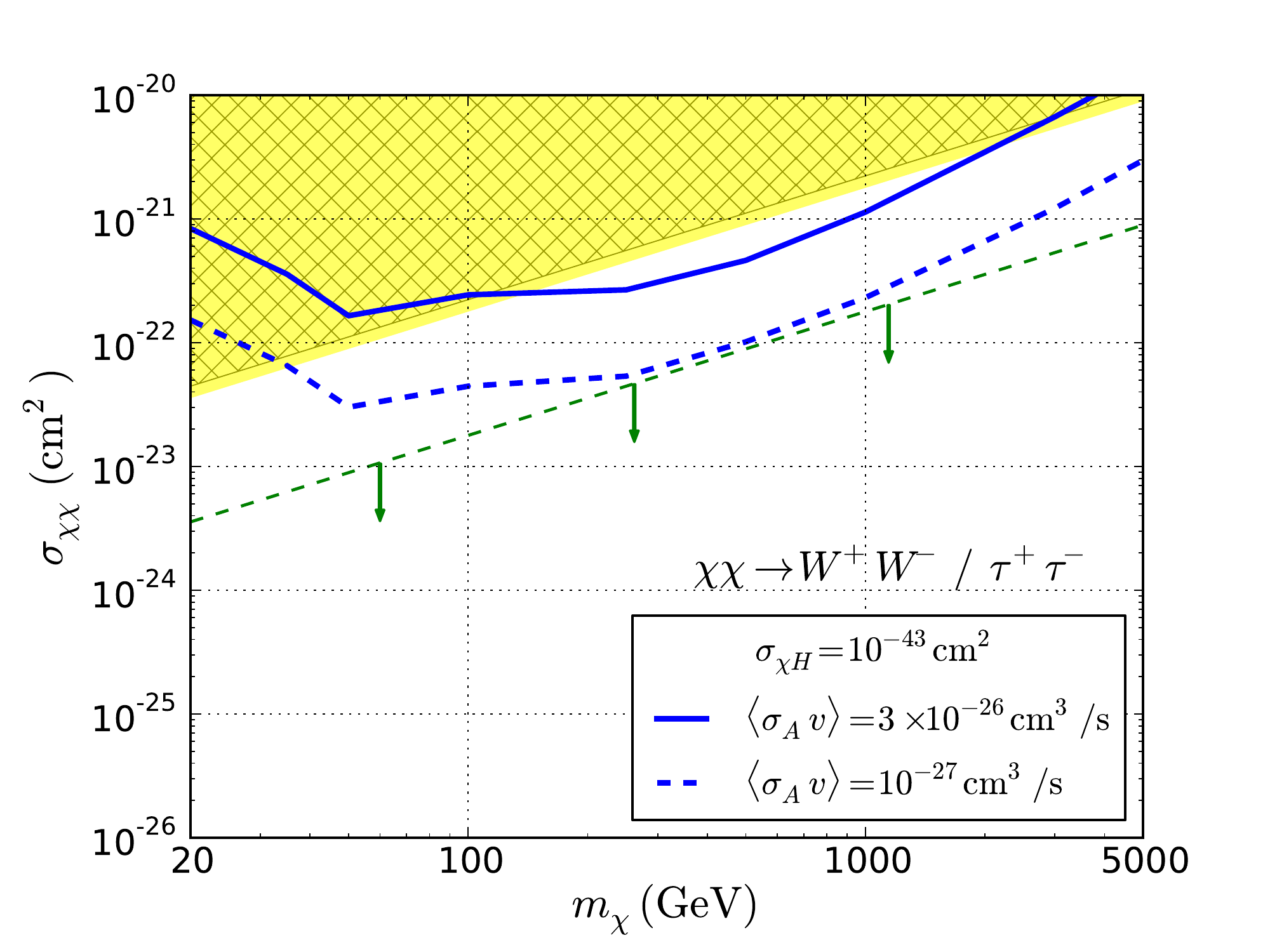}
\vspace*{-.15cm}
\caption{\label{fig:wwsd} Same as Figure~\ref{fig:wwexc} but now
  considering $\sigma_{\chi H}$ spin-dependent interactions. There are
no spin-dependent direct detection probes of the region shown in these plots.}
\end{center}
\end{figure}

\begin{figure}[t]
\begin{center}
\vspace*{-2.cm}
\hspace*{-1.2cm}
\includegraphics[scale=0.4]{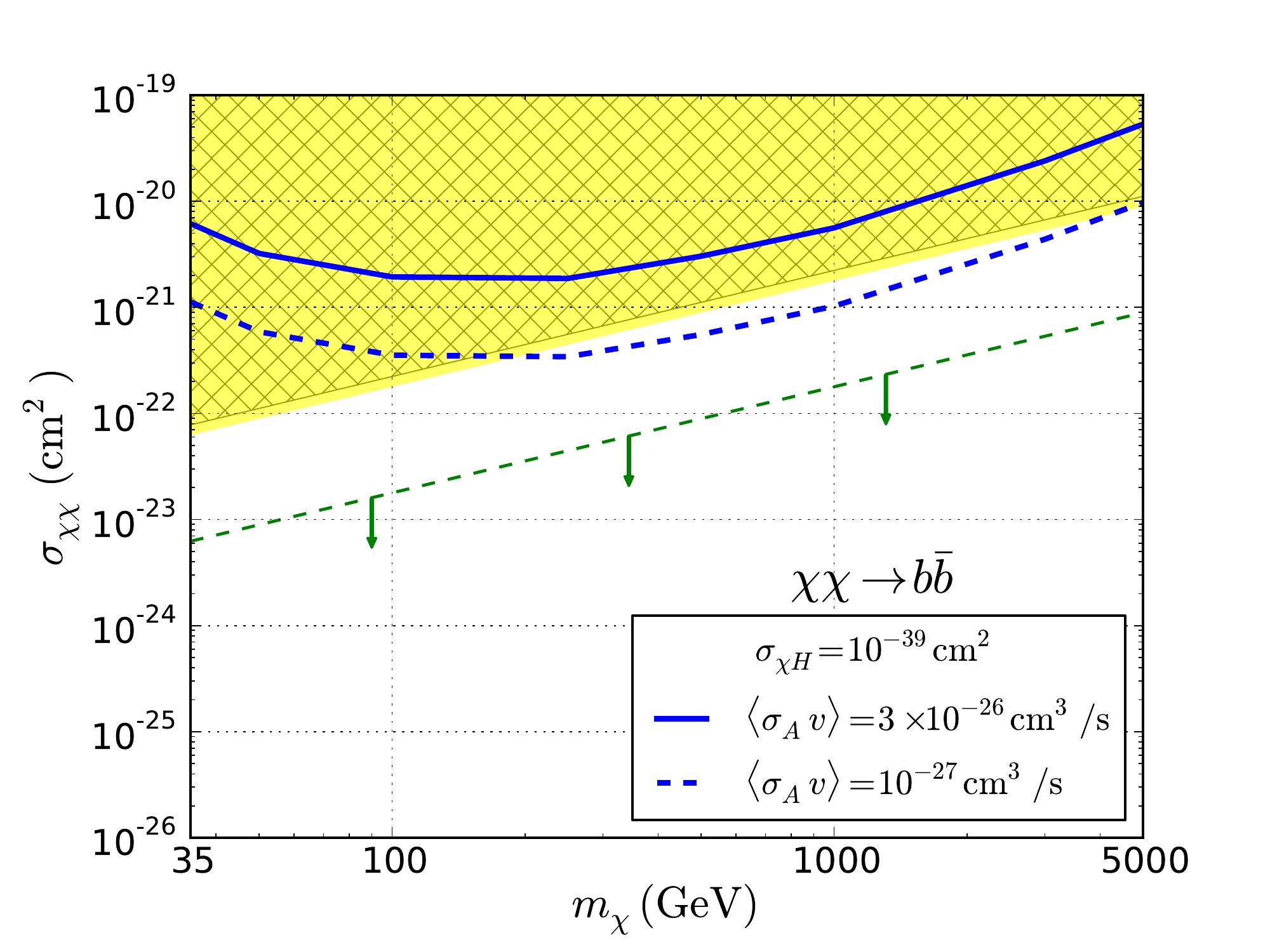}
\includegraphics[scale=0.4]{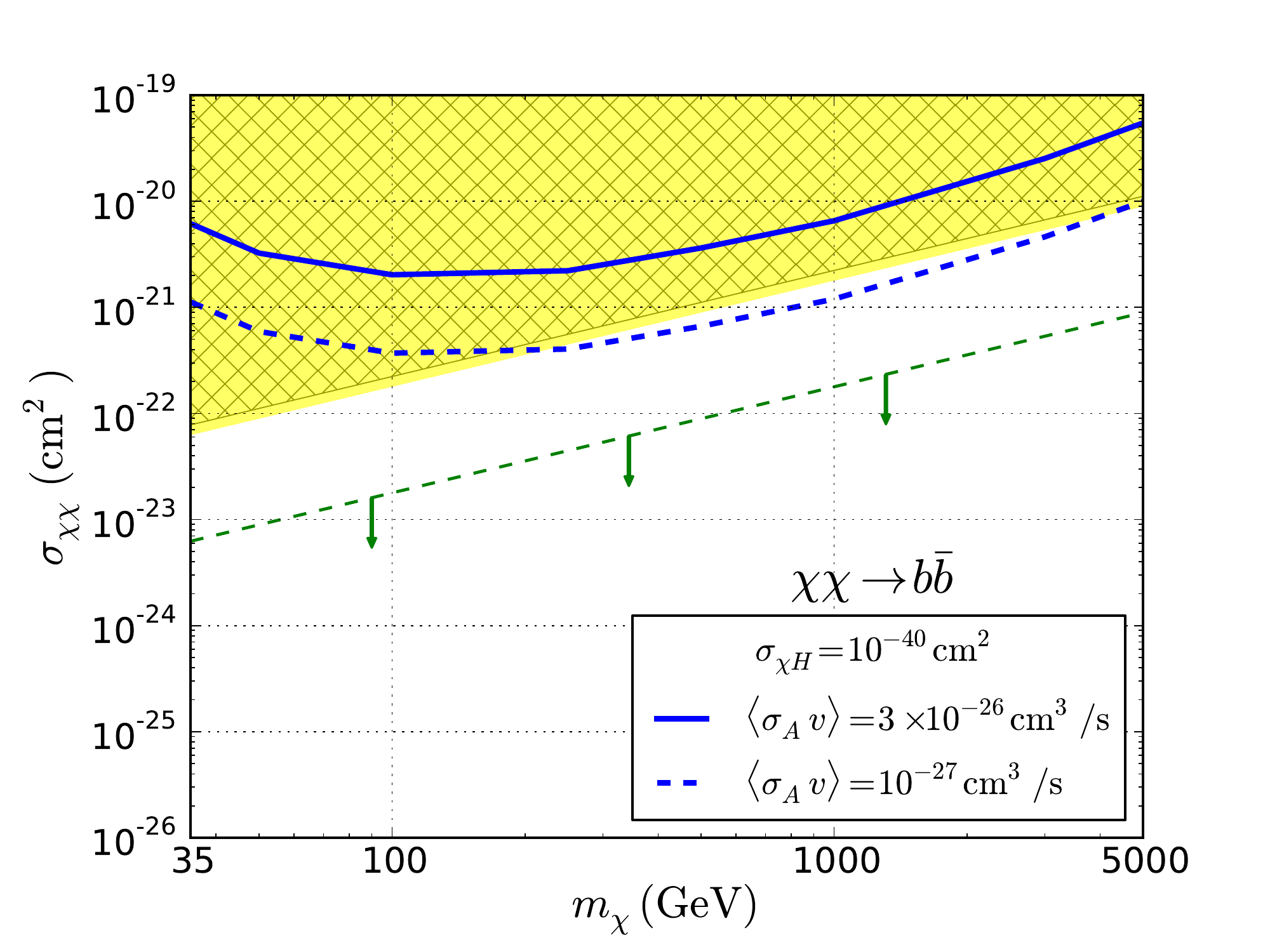}
\hspace*{-1.2cm}
\includegraphics[scale=0.4]{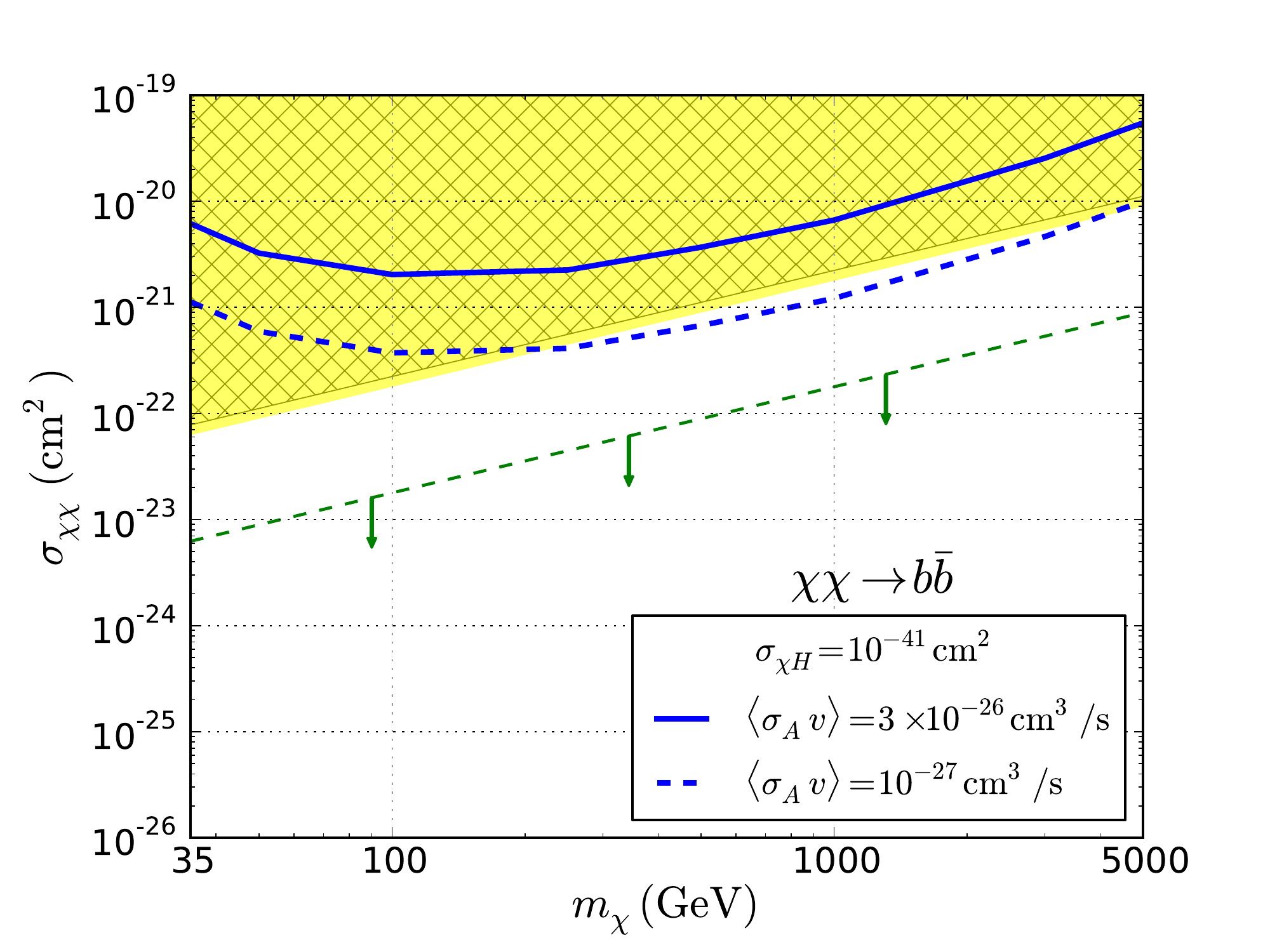}
\includegraphics[scale=0.4]{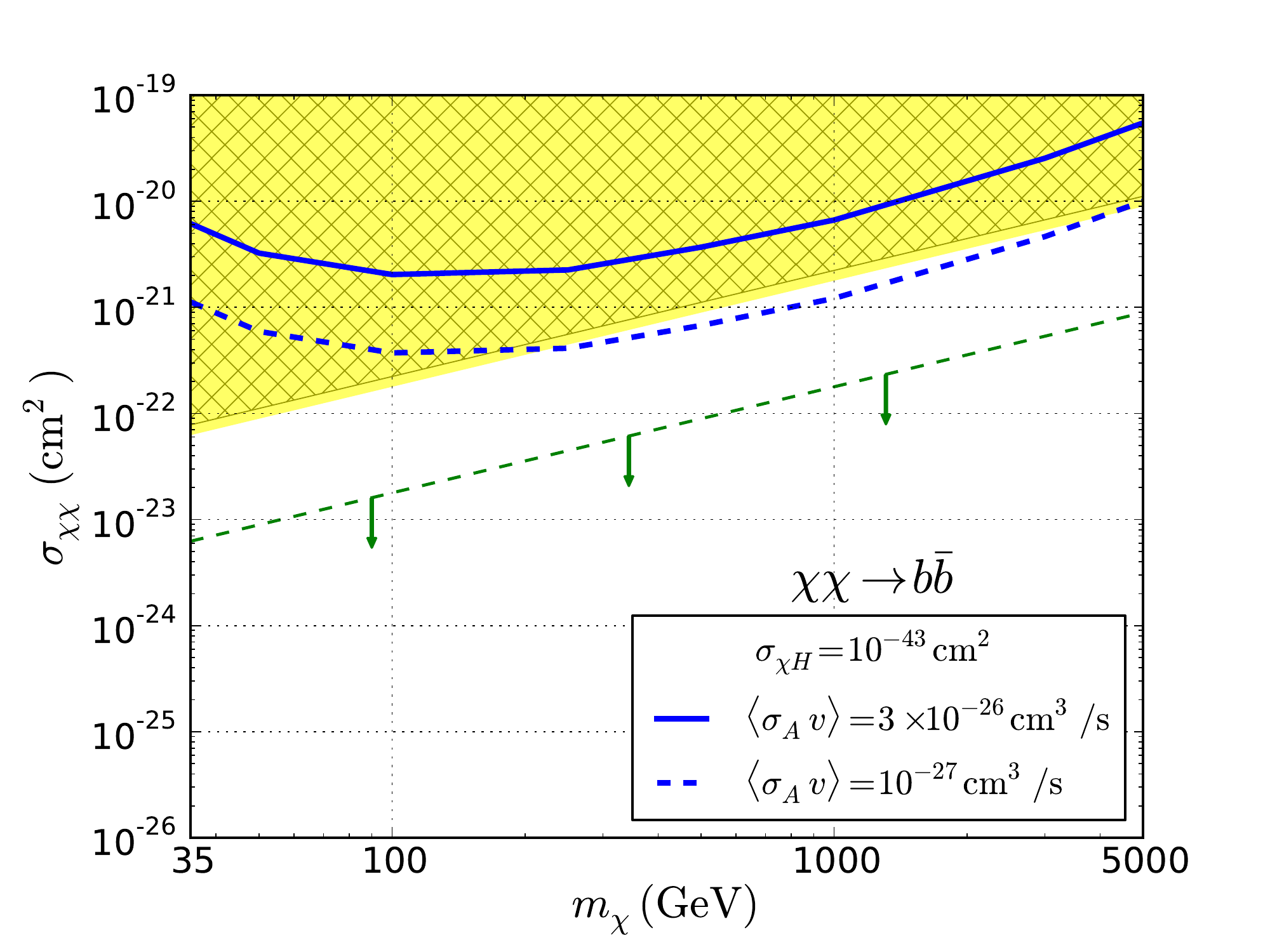}
\vspace*{-.15cm}
\caption{\label{fig:bbsd} Same as Figure~\ref{fig:bbexc} but now
  considering $\sigma_{\chi H}$ spin-dependent interactions.}
\end{center}
\end{figure}

We have also estimated the event rate for spin-dependent dark matter
interactions, which mainly consists of dark matter interacting with hydrogen in
the Sun. For the channels we are analyzing, IceCube has the most
stringent limits on the spin-dependent WIMP-proton cross section \cite{icprl}. We proceed in the same
way as for spin-independent scattering. Equation~\ref{eq:nev} depends
on the annihilation rate which in its turn depends on the capture
rate. For spin-dependent interactions the latter is given by equation
\ref{eq:gcsd} instead of equation \ref{eq:cpt}. Our results for the
$W^+W^-$ annihilation channel are shown in Figure \ref{fig:wwsd} and for
the $b\overline{b}$ channel  in Figure \ref{fig:bbsd}. We choose $\sigma_{\chi
  H}$ values that are not constrained by IceCube~\cite{icprl}.

Our spin-dependent exclusion region is larger than the spin
independent, which is expected since we assume larger cross sections than the ones for the spin-independent analysis. As for
the spin-independent, the exclusion region for the $b\overline{b}$ is
as expected smaller than the one for the $W^+W^- / \tau^+\tau^-$ channel.

\section{\label{conc} \bf Conclusions}

We have demonstrated that most  SIDM models with strong self-interacting cross
sections, at $\sigma_{\chi \chi} \gae \vartheta(10^{-22})$~cm$^2$
  (or $\vartheta(10^{-23})$)
  for $\left< \sigma_A v \right> = 3 \times 10^{-26}$~cm$^3$/s ($1 \times 10^{-27}$~cm$^3$/s), are ruled out if they annihilate into $W^+W^-$s. This
exclusion comes from the comparison of our predicted neutrino signal in the IceCube detector to their
observations~\cite{icprl}. This result is valid for both
spin-dependent and independent interactions, with the first one being
more stringent. Our results are summarized in figures \ref{fig:wwexc}
to \ref{fig:bbsd}.

If the assumptions made in the analyses presented in \cite{bullet,irv2} and \cite{zav}
  are correct, and dark
  matter annihilates mainly into the $W^+W^- / \tau^+\tau^-$ channel,
most of the significant SIDM scenarios are excluded
for thermal annihilation  cross sections $\left<\sigma_A v \right>
  \leq 10^{-27}$~cm$^2$. Under
these assumptions, all self-interacting models with $300 \lae
  m_\chi \lae 1000$~GeV,  and in the region identified by the dwarf spheroidals
    analysis as the one which would alleviate the CDM small scale
    potential problems, are now excluded. In this case, solutions to these
    problems will have to be encountered in different SIDM
scenarios, where, for example, the annihilation channel produces
  lower energy neutrinos, as for instance the $b\overline{b}$ channel. Another
  possibility is to consider that the self scattering
$\sigma_{\chi\chi}$ is velocity dependent \cite{zurek}. In our
  analysis, we determine the self scattering rate by considering a
  velocity independent $\sigma_{\chi\chi}$, as can be seen from
  equation~\ref{eq:gcc}. It is interesting to check the modifications
  to our results from a velocity dependent analysis~\cite{wip}.

 In relation to the $b\overline{b}$ channel, we independently
  confirm the Bullet cluster~\cite{bullet} and halo shapes~\cite{irv2}
  analyses. For $\left<\sigma_A v
  \right> \leq  \, 10^{-27}$~cm$^2$  a large region
  of strong $\sigma_{\chi \chi}$ is ruled out, and for $\left<\sigma_A v
  \right> = 3 \times 10^{-26}$~cm$^2$ most of the $\sigma_{\chi \chi}
~\protect\gae 5\times 10^{-21}$~cm$^2$ region is excluded.

 We also compared our results to the most stringent direct
  detection results, from the LUX \cite{lux} collaboration. As can be
  seen from Figures~\ref{fig:wwexc} and \ref{fig:bbexc}, direct
  detection experiments have not yet probed the region with
  $\sigma_{\chi n} \protect\lae 10^{-45}$~cm$^2$ which is probed by our analysis. 

\acknowledgments 
IA acknowledges the partial support from the Brazilian National Counsel for
Scientific Research (CNPq), and from the
European Union FP7  ITN INVISIBLES (Marie Curie Actions, PITN-
GA-2011- 289442). DSR was funded by the State of S\~{a}o Paulo
Research Foundation (FAPESP). CPH is partially supported by a grant from the Swedish Research Council.

\end{document}